\documentclass[pdflatex,sn-nature]{sn-jnl}

\usepackage{graphicx}%
\usepackage{multirow}%
\usepackage{amsmath,amssymb,amsfonts}%
\usepackage{amsthm}%
\usepackage{mathrsfs}%
\usepackage[title]{appendix}%
\usepackage{xcolor}%
\usepackage{textcomp}%
\usepackage{manyfoot}%
\usepackage{booktabs}%
\usepackage{algorithm}%
\usepackage{algorithmicx}%
\usepackage{algpseudocode}%
\usepackage{listings}%
\usepackage{soul}
\usepackage{xr}
\theoremstyle{thmstyleone}%
\theoremstyle{thmstyletwo}%

\theoremstyle{thmstylethree}%

\usepackage{dcolumn}

\newcommand{\ket}[1]{\left| #1 \right\rangle}
\newcommand{\bra}[1]{\left\langle #1 \right|}

\newcommand{\kp}{\mathbf k}

\newcommand{\rt}{\right}
\newcommand{\lf}{\left}
\renewcommand\Re{\operatorname{Re}}
\renewcommand\Im{\operatorname{Im}}

\newcommand{\la}{\langle}
\newcommand{\ra}{\rangle}

\renewcommand{\k}{\mathbf k}

\newcommand{\br}{\mathbf r}

\newcommand{\eq}[1]{Eq.~(\ref{#1})}

\newcommand{\mueven}{\mu_{\text{even}}}
\newcommand{\muodd}{\mu_{\text{odd}}}

\begin{document}

\title{Laser-dressed partial density of states}

\author[1,2]{\fnm{Tatiana} \sur{Bezriadina}}\email{tatiana.bezriadina@uni-hamburg.de}

\author*[1,2,3]{\fnm{Daria} \sur{Popova-Gorelova}}\email{daria.gorelova@b-tu.de}

\affil*[1]{I. Institute for Theoretical Physics, the University of Hamburg, Notkestr. 9, 22607 Hamburg, Germany}

\affil[2]{The Hamburg Centre for Ultrafast Imaging (CUI), Luruper Chaussee 149, 22607 Hamburg, Germany}

\affil[3]{Institute of Physics, Brandenburg University of Technology Cottbus-Senftenberg, Erich-Weinert-Stra\ss e 1, 03046 Cottbus, Germany}

 
 

\date{\today}
\maketitle
\begin{abstract}


\section*{Abstract}

The manipulation of material properties by laser light holds great promise for the development of future technologies. However, the full picture of the electronic response to laser driving remains to be uncovered. We present a novel approach to reveal details of the electron dynamics of laser-dressed materials, which consists of calculating and analysing the time-dependent partial density of states (PDOS) of materials during their interaction with a driving electromagnetic field. We show that the laser-dressed PDOS provides information about the structure of the bonds that form the laser-dressed electron density, analogous to the information that a PDOS can provide about the electron structure in a field-free case. We illustrate how our method can provide insights into the electron dynamics of materials in a site- and orbital-selective manner with calculations for a laser-dressed wurtzite ZnO crystal. Our work provides an analytical tool for the interpretation of subcycle-resolved experiments on laser-dressed materials and for the development of strategies for optical manipulation of material properties.


\end{abstract}

\maketitle


\section{\label{sec:level1}Introduction}

Coupling matter to an optical pulse can lead to significant changes in its electronic structure, opening the way for the optical manipulation of material properties. Laser dressing can modify the band structure of a material - the concept behind Floquet engineering  \cite{basov2017,oka2019floquet,rudner2020band,RevModPhys93421,wang2013observation, mahmood2016selective,ito2023build,Choi2025}, - and can be used to control its topological properties \cite{sie2019,mciver2020,zhou2023pseudospin}. Laser pulses can induce phase transitions \cite{stojchevska2014ultrafast,naseska2021first,johnson2023} and superconducting states \cite{fausti2011light, mitrano2016possible, cantaluppi2018pressure, budden2021evidence}. In the context of Floquet engineering, it is interesting to observe quasi-equilibrium laser-dressed states that arise from dynamics averaged over many optical cycles \cite{basov2017,oka2019floquet,rudner2020band,RevModPhys93421} and, in the context of ultrafast optoelectronics, to study laser-driven electron dynamics in a time-resolved manner \cite{langer2020few, paasch2016sub, hanus2021light, kwon2016semimetallization, kwon2016phz, Sommer2016, Garg2016, Higuchi2017, Siegrist2019, Ossiander2022, Heide2024}.

Due to advances in the ultrafast science, it is now possible to track electron dynamics with a subcycle temporal resolution, i.e. in a time-resolved manner during the interaction with a driving electromagnetic field \cite{Mashiko2016,tr-XAS4,Popova-Gorelova_2024,Schultze_2014,Lucchini_2016,tr-XAS2,sidiropoulos2021probing,sidiropoulos2023enhanced}. A beneficial technique for probing electron dynamics is attosecond core-level x-ray absorption spectroscopy. This technique has been used to resolve electron, hole and phonon dynamics in materials with a subfemtosecond resolution including times during the interaction with a driving pulse \cite{Schultze_2014,Lucchini_2016,tr-XAS2,sidiropoulos2021probing,sidiropoulos2023enhanced}. Another successful technique for probing laser-dressed dynamics is time- and angle-resolved photoemission spectroscopy. Using this technique, Floquet states have been observed in momentum space \cite{wang2013observation,zhou2023pseudospin} and the evolution of emerging Floquet sidebands has been tracked over time \cite{mahmood2016selective,ito2023build, Choi2025}. These novel experiments capable of resolving laser-dressed electron dynamics during laser driving motivate theoretical developments and advances in ab initio computational tools to describe electron dynamics under the action of light  \cite{PhysRevB.93.045124,XOWM,PhysRevResearch.7.L022042,PhysRevB.102.115106,Wopperer2017,Sato2025,Picon_2019,doi:10.1021/acs.jctc.2c00674,popova2024atomic}.


One of the main theoretical tools to analyze properties of laser-dressed electronic systems is the laser-dressed density of states (DOS) \cite{oka2009photovoltaic,zhou2011optical}. The laser-dressed DOS describes the changes of electronic population due to the interaction with light. Analysis of its changes depending on the properties of a driving pulse offers a way to tune optical, topological and transport properties of laser-driven materials \cite{PhysRevLett.76.4576,enders2004electronic,lima2008effect,lima2009terahertz,oka2009photovoltaic,zhou2011optical,suarez2012radiation,calvo2012laser,PhysRevB.91.045439,PhysRevB.102.220301,PhysRevResearch.2.023274,Park2022,PhysRevB.108.235429,LIU2024415798,Zhan2024}. In a field-free case, a partial density of states (PDOS) is an extension of DOS, which provides its orbital decomposition. A PDOS is a valuable computational tool of characterizing the electronic structure of materials in a ground state that reveals the contribution of specific orbitals to the bond formation. For a nonequilibrium state, similar information is especially relevant in view of recent experiments that demonstrate optical control of structural dynamics and ultrafast, optically induced bond formation or breaking \cite{Jiang2024,Hutchison2023,doi:10.1126/sciadv.aar3867,Frigge2017,Kolobov2011,Horstmann2020}. Here, we introduce a laser-dressed PDOS that describes laser-induced changes in the electronic population in a site- and orbital-resolved manner.  

Analyzing a laser-dressed DOS, it is common to focus only on its time-averaged part. Here, we also go beyond this standard approach and analyze both the time-averaged and the time-dependent part of the laser-dressed PDOS. We show that the time-dependent part of the PDOS reveals the contribution of specific orbitals to the time-dependent changes in the electron density. This information provides a structure of instantaneous bonds formed during the interaction with light. On the one hand, it provides new details about microscopic optically-induced dynamics, and, on the other hand, it can be useful for the interpretation of subcycle-resolved measurements. We illustrate our approach with the calculations of the PDOS of a laser-dressed wurtzite ZnO crystal.

Our calculations of the PDOS are based on an ab initio computational framework within the Floquet-Bloch formalism. The Floquet-Bloch formalism is a powerful method to describe laser-dressed electron dynamics beyond the perturbation theory \cite{hsu2006floquet, faisal1997floquet, tzoar1975theory}.The Floquet formalism implies periodic driving. This assumption is justified for a strong optical field comprising 10--15 cycles \cite{Lucchini2022,PhysRevB.104.L060303}. Our study is thus relevant for optical manipulation by laser pulses of several tens femtoseconds or longer. We combine the Floquet-Bloch formalism with the density functional theory (DFT) to obtain laser-dressed states. Laser-dressed states provide observables of a laser-dressed electronic system such as a time-dependent electronic density or an electron current density \cite{hsu2006floquet,XOWM,popova2024microscopic}. Here, we use the DFT in the combination with the linearized augmented plane wave plus localized orbitals (LAPW$+$lo) method \cite{0} to obtain laser-dressed states and use them to calculate PDOS. We select the LAPW$+$lo method, since it provides a full potential treatment of an electronic structure, and is convenient to project the density of states on particular orbitals.

\section{Results}

\subsection{Derivation of a partial density of states}

Let us first consider a momentum- and angle-resolved PDOS of a periodic material in the ground state. In the following, we refer to the momentum- and angle-resolved PDOS as simply PDOS for convenience. A PDOS within LAPW+lo basis set can be defined via the localized part of a Kohn-Sham orbital inside a muffin-tin (MT) region. This part is given by
\begin{align}
\label{eq:wf}
\varphi^{\rm MT}_{n\mathbf{k}}(\mathbf{r}_{\alpha})=\sum\limits_{lm}F^{n\mathbf{k}}_{lm\alpha}(r_{\alpha})Y_{lm}(\hat{\mathbf{r}}_{\alpha}),
\end{align}
where the radius vector $\mathbf{r}_{\alpha}$ is defined inside a MT sphere of radius $R_\alpha$ centered on atoms labeled $\alpha$, $n$ is a band and spin index, and $\kp$ is the Bloch wave vector \cite{SJOSTEDT}. $Y_{lm}(\hat{\mathbf{r}}_{\alpha})$ is a spherical harmonic function of degree $l$ and order $m$. The function $F^{n\mathbf{k}}_{lm\alpha}(r_{\alpha})$ depends on a material and is determined during electron-structure calculations. We transform the basis to real spherical harmonics, $\mathcal{Y}_{lm}$ \cite{orbitals}, and express $F^{n\mathbf{k}}_{lm\alpha}(r_{\alpha})$ in the real harmonics basis via 
\begin{align}
\mathcal{F}^{n\mathbf{k}}_{lm\alpha} =
\begin{cases}
\frac{i}{\sqrt{2}}((-1)^{m}F^{n\mathbf{k}}_{lm\alpha}-F^{n\mathbf{k}}_{l,-m\alpha}), \text{if} \;\; m<0,\\
F^{n\mathbf{k}}_{lm\alpha}, \text{if} \;\; m=0,\\
\frac{1}{\sqrt{2}}((-1)^{m}F^{n\mathbf{k}}_{lm\alpha}+F^{n\mathbf{k}}_{l,-m\alpha}), \text{if} \;\; m>0.
\end{cases}
\label{eq:realbasis}
\end{align}
Analogously to a definition of a DOS, we define an atomic-resolved DOS as $\sum_{n,\kp}\int d^3 r|\varphi^{\rm MT}_{n\mathbf{k}}(\mathbf{r}_{\alpha})|^2\delta(\omega-\epsilon_{n,\kp})$ and decompose it in terms of different $l$ and $m$ components to obtain a PDOS as 
\begin{align}
{D}^{\alpha}_{lm}(\omega) = \sum_{n\mathbf{k}} \int_{0}^{R_{\alpha}} dr_{\alpha}\, r_{\alpha}^2|\mathcal F^{n\mathbf{k}}_{lm\alpha}(r_{\alpha})|^2  \delta (\omega - \epsilon_{n\mathbf{k}}),
\end{align}
where the sum is taken over all Kohn-Sham orbitals $ \ket{\varphi_{n\mathbf{k}}} $ with eigenenergies $ \epsilon_{n\mathbf{k}} $. Here, indices $l$ and $m$ refer to specific angular characteristics rather than fixed quantum numbers $l$ and $m$. For example, the component with $l=1$ and $m=+1$ corresponds to the $p_x$-projected PDOS , $l=1$ and $m=-1$ -- to the $p_y$-projected PDOS, and $l=1$ and $m=0$ -- to the $p_z$-projected PDOS.

We now consider a periodic material exposed to a spatially uniform and temporally periodic electromagnetic field with the electric field $\mathbf{E}(t)=\mathbf{F}_0 \sin\Omega t$. Here, $\mathbf{F}_0$ is the amplitude of the field. The Hamiltonian of the irradiated system is time-periodic, $H_{\rm el-em}(t+T)=H_{\rm el-em}(t)$, with the period $T = 2\pi/\Omega$ and is given by
\begin{equation}
H_{\rm el-em}(t) = \frac{1}{2}\left(\mathbf{p}+\frac{1}{c}\mathbf{A}(t)\right)^2 + V_{\rm c}(\mathbf{r}),
\end{equation}
where $V_{\rm c}(\mathbf{r})$ is a space-periodic crystal potential, $\mathbf{p}$ is the momentum of an electron, $\mathbf{A}(t)$ is the vector potential, and $c$ is the speed of light.
We use atomic units, in which $m_{\rm e}=|e|=\hbar=1$, for this and the following expressions. We assume the dipole approximation.

We apply the Floquet-Bloch formalism \cite{hsu2006floquet,faisal1997floquet,tzoar1975theory,Shirley,multiphoton,XOWM} to describe the interaction of a material with the electromagnetic field beyond a perturbative regime. Within this formalism, a solution of the time-dependent Schr\"odinger equation for a spatially periodic system exposed to a periodic driving is a Floquet-Bloch state given by
\begin{equation}
\label{eq:FB}
\Psi_{i\mathbf{k}}(\mathbf{r},t) = \frac{1}{\sqrt{V}}e^{-iE_{i\mathbf{k}}t}\sum_\mu\sum\limits_{n}C^{i}_{n\mathbf{k}\mu}e^{-i\mu\Omega t}\varphi_{n\mathbf{k}}(\mathbf{r}),
\end{equation}
where $i$ is an index and $E_{i\mathbf{k}}$ is the corresponding quasienergy of the Floquet-Bloch state, $\mu$ is an integer, and the wave function is normalized to the system volume $V$. The coefficients $C^{i}_{n\mathbf{k}\mu}$ are determined by diagonalization of a Floquet-Bloch Hamiltonian.

A DOS of a periodically-driven electronic system as derived in Ref.~\cite{zhou2011optical} is
\begin{equation}
\label{eq:1DOS}
D(\omega,t) = \frac{1}{2\pi}\sum\limits_{\mathbf{k}}\mathrm{Tr}\hat{A}_{\mathbf{k}}(\omega,t),
\end{equation}
where $\hat{A}_{\mathbf{k}}(\omega,t)$ is a spectral function that has the form
\begin{equation}
\label{eq:SpecFun}
\hat{A}_{\mathbf{k}}(\omega,t) = \int\limits_{-\infty}^{\infty}d\tau e^{i\omega\tau}\sum\limits_{i}\ket{\Psi_{i\mathbf{k}}\left(t+\frac{\tau}{2}\right)}\bra{\Psi_{i\mathbf{k}}\left(t-\frac{\tau}{2}\right)}
\end{equation}
in the non-interacting limit. Analogously to the derivation of a PDOS in a field-free case, we determine a laser-dressed PDOS by substituting $\varphi^{\rm MT}_{n\mathbf{k}}(\mathbf{r}_{\alpha})$ instead of $\varphi_{n\mathbf{k}}$ in the expression for a Floquet-Bloch state in Eq.~\eqref{eq:FB}. We substitute the resulting spatially-localized part of  $\Psi_{i\mathbf{k}}$ into Eqs.~\eqref{eq:1DOS} and \eqref{eq:SpecFun} and obtain a laser-dressed PDOS
\begin{align}
\mathcal D^\alpha_{lm}(\omega,t) = \la \mathcal D^\alpha_{lm}(\omega) \ra+2\sum\limits_{\mu\geq 1}\Re\lf[e^{i\mu\Omega t}\tilde{D}^{\alpha(\mu)}_{lm}(\omega)\rt]
\label{DOS}
\end{align}
with the amplitudes
	\begin{align}
	\tilde{D}^{\alpha(\mu)}_{lm}(\omega) =   \sum\limits_{i\mathbf{k}}\sum\limits_{n,\mu'} C^{i*}_{n\mathbf{k}\mu'+\mu}C^{i}_{n\mathbf{k}\mu'}\int\limits_{0}^{R_{\alpha}}dr_{\alpha}r^2_{\alpha}|\mathcal F^{n\mathbf{k}}_{lm\alpha}(r_{\alpha})|^2\delta\Big(\omega-\lf[E_{i\mathbf{k}}+(\mu'+\mu/2)\Omega\rt]\Big).
	\label{Eq_Dalm_Amp}
	\end{align}
The first term in Eq.~\eqref{DOS} is the zero-order amplitude $\la \mathcal D^\alpha_{lm}(\omega) \ra = \tilde{D}^\alpha_{\mu=0 lm}(\omega) $ that gives a PDOS averaged over the period $T$ of the driving electromagnetic field. It captures the steady-state behavior of a system under a continuous periodic excitation. The second term in Eq.~\eqref{DOS} is the time-dependent part of a PDOS.

Time-reversal symmetry determines a temporal behavior of a PDOS. If an electronic system with time-reversal symmetry is driven by a linearly polarized pulse, a property $C^{i}_{n\mathbf{k}\mu}=(-1)^{\mu}\lf (C^{i}_{n-\mathbf{k}\mu}\rt)^*$ derived in \cite{popova2024microscopic} applies. It is also true for a system with time-reversal symmetry that $F^{n\mathbf{k}}_{lm\alpha}(r_{\alpha})=(-1)^{m}\lf [F^{n-\mathbf{k}}_{l-m\alpha}(r_{\alpha})\rt]^*$. This relates the terms with a positive and negative $\kp$ in the sum in \eq{Eq_Dalm_Amp} as complex conjugates multiplied by a factor $(-1)^\mu$. The amplitudes of a PDOS are then real for even orders and are imaginary for odd orders, which leads to
\begin{align}
\mathcal{D}^{\alpha}_{lm}(\omega,t) = \la \mathcal D^\alpha_{lm}(\omega) \ra &- \sum\limits_{\mu_{\rm odd}\geq 1} \mathcal{D}^{\alpha(\muodd)}_{lm}(\omega)\sin(\mu_{\rm odd}\Omega t) \nonumber \\
&+ \sum\limits_{\mu_{\rm even}\geq 2}\mathcal{D}^{\alpha(\mueven)}_{lm}(\omega)\cos(\mu_{\rm even}\Omega t),
\label{Eq_damplitudes}
\end{align}
where $\mathcal{D}^{\alpha(\muodd)}_{lm}(\omega) = 2\Im\lf({\tilde{D}}^{\alpha(\muodd)}_{lm}(\omega)\rt)$ and $\mathcal{D}^{\alpha(\mueven)}_{lm}(\omega) = 2\Re\lf({\tilde{D}}^{\alpha(\mueven)}_{lm}(\omega)\rt)$ are real-valued amplitudes.


\subsection{Field-free and time-dependent PDOS of ZnO}

We illustrate how a laser-dressed PDOS can be used to analyze the behavior of a laser-driven material with calculations for a wurtzite ZnO crystal. The first observation of high harmonic generation in a bulk crystalline solid has been reported in a wurtzite ZnO crystal , showing a distinctly non-perturbative harmonic spectrum under a 3.25 $\mu$m driving laser at intensities exceeding 1 $\rm TW/cm^2$. We consider the driving electromagnetic field with a photon energy of 1.55 eV, an intensity of $5 \times 10^{12} \, \rm W/cm^2$, and the electric field polarization $\mathbf{\epsilon}$ aligned along the $z$ axis. High harmonic generation in ZnO was achieved by pulses of similar intensities and with a photon energy below the band gap \cite{HHG1}. Thus, interaction of ZnO with the field should lead to nonlinear and multiphoton effects. 

\begin{figure}[tb]
	\includegraphics[width=\linewidth]{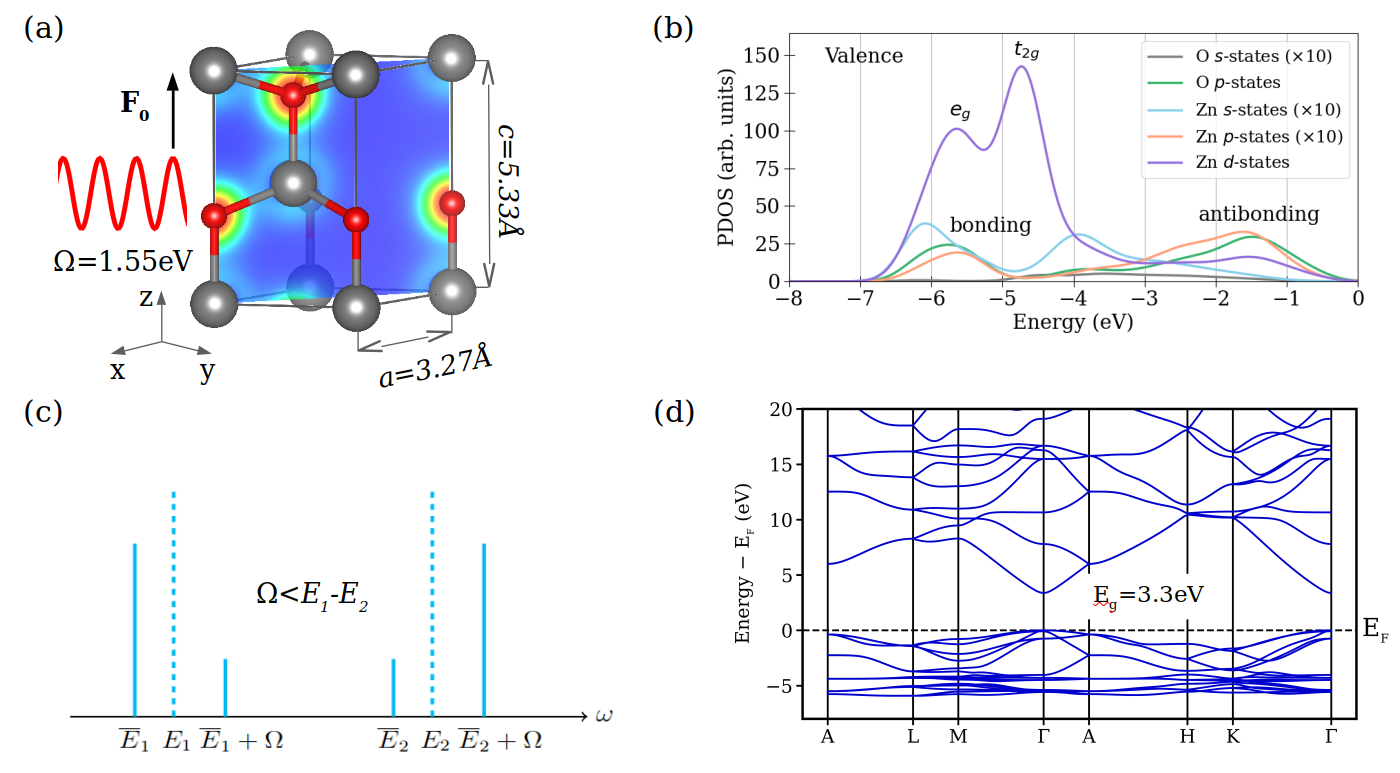}
	\caption{ (a) Structure of a wurtzite ZnO, where the red spheres represent O and the gray spheres represent Zn, and a cut through the electron density. This and the following densities are visualized using VESTA \cite{momma2011vesta}. (b) PDOS of a wurtzite ZnO in the ground state. The Zn $s$-, $p$- and O $s$-projected PDOS are magnified by a factor of ten. The Fermi energy, $E_\mathrm{F}$, is set to 0 eV. (c) Illustration of the DOS of the two-level laser-driven system. The laser-dressed DOS is shown in solid lines next to the DOS in the absence of laser driving in dotted lines. (d) Band structure of a wurtzite ZnO in the ground state.}
	\label{fig:DOS}
\end{figure}

We first discuss the PDOS of wurtzite ZnO in the ground state. The calculations and detailed analysis of the PDOS of ZnO have been reported in Ref.~\cite{ZnO}, which we now briefly review.  Fig.~\ref{fig:DOS}(b) shows the calculated PDOS of the valence states, which is in good agreement with the calculations in Ref.~\cite{ZnO}. The valence region is dominated by electronic states of strong O $2p$ and Zn $3d$ character. The Zn $s$-, $p$- and O $s$-projected PDOS are magnified by a factor of ten. The tetrahedral environment around the Zn atom leads to the $e_g-t_{2g}$ splitting of the $d$ states at energies around -5 eV -- -6 eV below the Fermi level. The $e_g$ states hybridize with the O $2p$ states to form bonding and antibonding states. 


\begin{figure*}[htbp]
\includegraphics[width=0.99\linewidth]{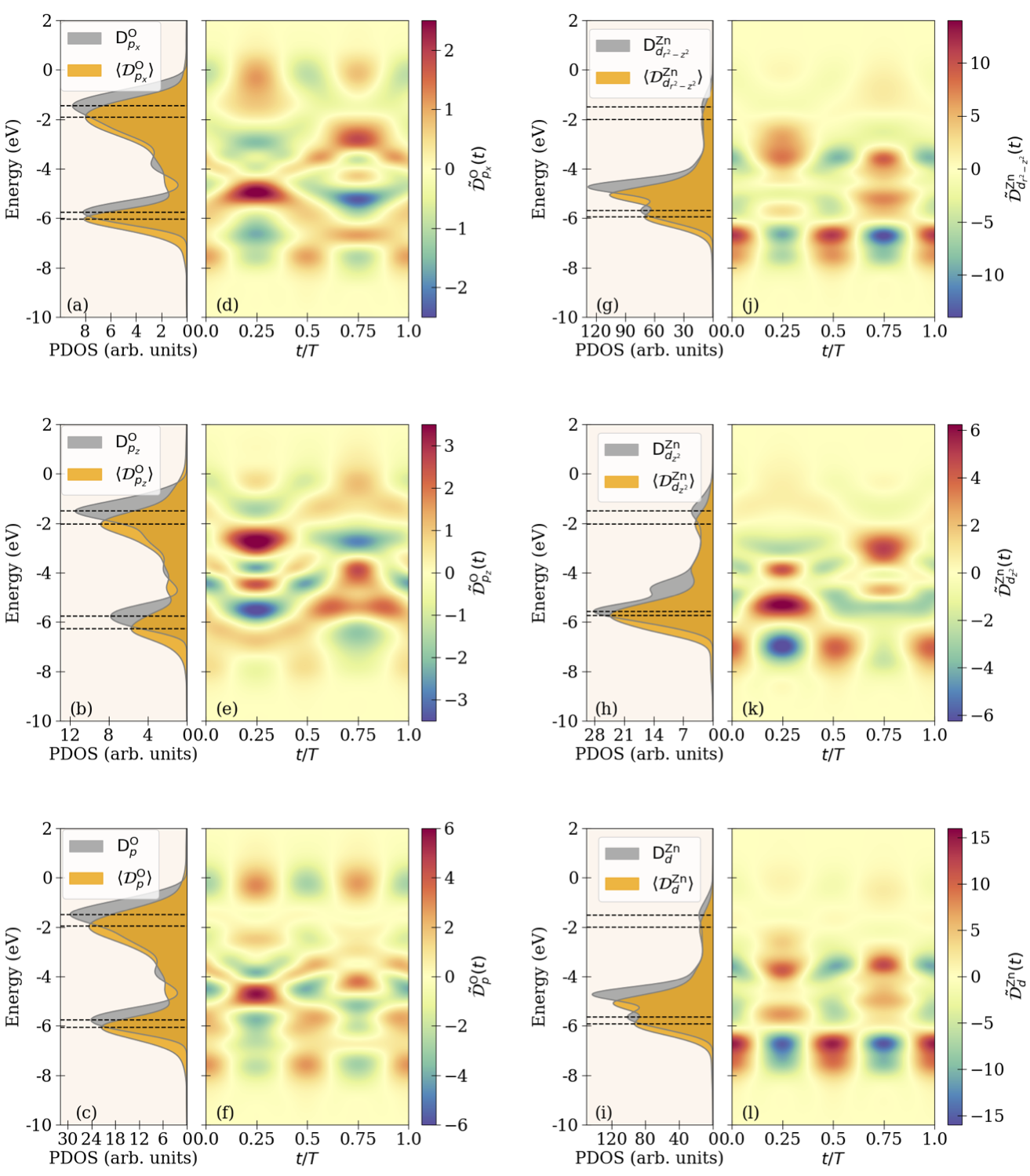}
	\caption{Angle-resolved and -unresolved O $p$- and Zn $d$-projected partial density of states (PDOS). The field-free PDOS is shown in gray and the time-averaged laser-dressed PDOS is shown in orange for the (a) O $p_x$-projected PDOS; (b) O $p_z$-projected PDOS; (c) O $p$-projected PDOS. The time-dependent part of the laser-dressed PDOS for the (d) O $p_x$-projected; (e) O $p_z$-projected PDOS; (f) O $p$-projected PDOS. (g)-(i) The same as (a)-(c), but for the (g) Zn $d_{r^2-z^2}$-projected PDOS, which is the PDOS projected on the Zn $d$-orbitals with an exception of the $d_{z^2}$ orbital; (h)  Zn $d_{z^2}$-projected PDOS; (i) Zn $d$-projected PDOS. (j)-(l) The same as (d)-(f), but for the (j) Zn $d_{r^2-z^2}$-projected PDOS; (k) Zn $d_{z^2}$-projected PDOS; (l) Zn $d$-projected PDOS.}
	\label{Fig_LDPDOS}
\end{figure*}

We now turn to the laser-dressed PDOS of ZnO, which we show in Fig.~\ref{Fig_LDPDOS}. We show the PDOS of the states that have emerged from the valence states shown in Fig.~\ref{fig:DOS}(b). Under the assumption of adiabatic switching, the Floquet-Bloch states that have a maximum overlap with the valence electronic state are occupied. Throughout the article, we analyze the corresponding PDOS of these states.  Each set of plots in Fig.~\ref{Fig_LDPDOS} shows a time-averaged PDOS next to a corresponding field-free PDOS in the left panels and the time-dependent part of the PDOS in the right panels. The total laser-dressed PDOS, which is the sum of the time-averaged and time-dependent parts, is positive for all energies.  We assume that the driving electromagnetic field is polarized along the $z$ axis. Therefore, we consider separately the PDOS projected on the oxygen $p_z$ orbitals aligned along the electric field and the oxygen $p_x$ orbitals aligned perpendicular to the electric field. The former PDOS is identical to the PDOS projected on the oxygen $p_y$ orbitals. We also consider separately the PDOS projected on the zinc $d_{z^2}$ orbitals and the sum of the PDOS projected on the remaining zinc $d$ orbitals, which we denote as $\mathrm D^{\rm Zn}_{d_{r^2-z^2}}$ for the field-free case and $\mathcal D^{\rm Zn}_{d_{r^2-z^2}}$ for the laser-dressed case. Note that our choice of axes is different from the conventional choice of axes in the crystal field theory that explains the $e_g-t_{2g}$ splitting. With our choice of axes, the $d_{z^2}$ orbital contributes to both the $e_g$ and $t_{2g}$ peaks.

\subsection{Time-averaged PDOS}

Our first observation is that all time-averaged PDOSs below the ground-state Fermi level shift toward higher binding energies. The dotted lines in Fig.~\ref{Fig_LDPDOS} go through the maxima of the laser-dressed and field-free PDOS. They highlight that the values of energy shifts vary depending on a binding energy and a projection.


To facilitate the understanding of the time-averaged laser-dressed PDOS, we revisit the laser-driven two-level system described within the Jaynes-Cummings model \cite{Jaynes-Cummings}. Let us consider a system with the energies $E_{1}$ and $E_2$ and corresponding wave functions $\phi_1(\br)$ and $\phi_2(\br)$ driven by a single-mode electromagnetic field with the frequency $\Omega<E_2-E_1$. The system of the matter and the light has states with the energies that are different from $E_1$ and $E_2$. This phenomenon is the optical Stark effect \cite{drake2007springer}. The shifted energy levels $\overline{E}_{1,2}$ obtained in the rotating wave approximation and taking only one-photon absorption processes into account are given by \cite{Jaynes-Cummings}:
\begin{align}
&\overline{E}_{1} =E_1 -  \frac{\sqrt{\Delta^2 + \Omega_R^2} -\Delta}{2},\nonumber\\
&\overline{E}_{2} = E_2  + \frac{\sqrt{\Delta^2 + \Omega_R^2}-\Delta }{2}-\Omega.
\label{eq:energy}
\end{align}
Here, $\Delta=E_2-E_1-\Omega $ is the detuning and $\Omega_R$ is the Rabi frequency. The corresponding eigenstates are given by\cite{Jaynes-Cummings}:
\begin{align}
&\ket{\overline\Psi_1} = \cos\Theta\ket{1,N} - \sin\Theta\ket{2,N-1},\label{Eq_eigenstates}\\
&\ket{\overline\Psi_2} = \sin\Theta\ket{1,N} + \cos\Theta\ket{2,N-1}.\nonumber
\end{align}
Here, the basis states $\ket{i,N-\mu}$ are product states formed by the electronic states $|i=1\ra$ and $|i=2\ra$, and Fock states of photons $|N-\mu\ra$, where $N-\mu$ is the number of photons in the mode of the considered single-mode field and $N\gg \mu$ is the number of the photons before the interaction with the electronic system. The coefficients in front of the basis states depend on the ratio between the detuning and the Rabi frequency via $\Theta = \arctan\lf(\Omega_R/\Delta\rt)/2$.

The connection of a laser-dressed picture to the Floquet picture has been discussed in Ref.~\cite{Shirley}. The corresponding Floquet states (see \eq{eq:FB}) are
\begin{align}
{\Psi_1}(\br,t) = \frac{e^{-i\overline E_{1}t}}{\sqrt{V}}\lf(\cos\Theta \phi_1(\br) - \sin\Theta e^{-i\Omega t} \phi_2(\br)\rt),\label{Eq_Floquet_twolevel}\\
{\Psi_2}(\br,t) = \frac{e^{-i\overline E_{2}t}}{\sqrt{V}}\lf(\sin\Theta \phi_1(\br) +  \cos\Theta e^{-i\Omega t} \phi_2(\br)\rt).\nonumber
\end{align}
The density of states according to \eq{eq:1DOS} is
\begin{align}
D(\omega) =& \cos^2\Theta\delta(\omega - \overline E_1)+\sin^2\Theta\delta\lf(\omega - \lf[\overline E_1+\Omega\rt]\rt)\label{eq_DOS2level}\\
&+ \sin^2\Theta\delta\lf(\omega - \overline E_2\rt)+\cos^2\Theta\delta\lf(\omega - \lf[\overline E_2+\Omega\rt]\rt)\nonumber
\end{align}
and is shown in Fig.~\ref{fig:DOS}(c).

If $\Omega_R$ is relatively small, then $\cos\Theta$ is noticeably larger than $\sin\Theta$. In this case, we can identify the state $\ket{\Psi_1}$ as the state that has emerged from the state $|1\ra$, and the state $\ket{\Psi_2}$ as the state that has emerged from the state $|2\ra$. The first two peaks of the DOS in \eq{eq_DOS2level} are then due to the state $|\Psi_1\ra$ that has emerged from the ground state $|1\ra$ and the last two peaks are due to the state $|\Psi_2\ra$ that has emerged from the excited state.  The DOS of the field-free two level system consists of two delta peaks $\delta(\omega -  E_1)+\delta(\omega - E_2)$. With \eq{eq_DOS2level}, we obtain that these peaks are split into four peaks in the laser-dressed case, see Fig.~\ref{fig:DOS}(c). Two of these peaks would be the main intense peaks weighted by $\cos^2\Theta$ centered at the new energies $\overline E_1$ and $\overline E_2+\Omega$, and each main peak would have a side peak weighted by $\sin^2\Theta$.  

The intensity of the peaks in the laser-dressed DOS in \eq{eq_DOS2level} gives the probability of occupation of a particular electronic state. For example, the first two peaks in \eq{eq_DOS2level} are due to the eigenstate $|\Psi_1\ra$. In this state, the occupation probability of the electronic state $|1\ra$ is equal to $\cos^2\Theta$, and the occupation probability of the electronic state $|2\ra$ is $\sin^2\Theta$. The nature of the side peak shifted by $\Omega$ can then be intuitively understood as an indication that the state $|2\ra$ can be excited due to absorption of a photon with an energy $\Omega$.



Let us consider the intensive peaks in the PDOS in Fig.~\ref{Fig_LDPDOS}. Their position is blue-shifted relative to the corresponding positions in the field-free distributions, and the intensity of the peaks is reduced. We also observe that the distributions become broader than the field-free distributions. Although the two-level system does not account for nonlinear and multi-photon effects, it can provide some understanding of the notable changes in the PDOS. In Fig.~\ref{fig:DOS}(c), the peak centered at $E_1$ in the field-free DOS blue-shifts to the new energy position centered at $\overline E_1$ and its intensity decreases from 1 to $\cos^2\Theta$. Analogously, the intensive peaks in the PDOS in Fig.~\ref{Fig_LDPDOS} blue-shift and decrease. And the appearance of  side peaks lead to the broadening of the distribution.


The antibonding peaks at about -1.5 eV in all field-free PDOS shown in Fig.~\ref{Fig_LDPDOS} are synchronously shifted by about 0.5 eV in the corresponding time-averaged PDOS. The bonding peaks at about -6 eV in the PDOS projected along the orbitals not aligned parallel to the electric field, namely O $p_x$ and Zn $d_{r^2-z^2}$, are also synchronously shifted by the same value of 0.2 eV (see Fig.~\ref{Fig_LDPDOS}(a) and (g)). However, the shift of the bonding peak is different in the PDOS projected along the orbitals aligned along the electric field (see Fig.~\ref{Fig_LDPDOS}(b) and (h)). This misalignment of the bonding peaks between the O $p$ and Zn $d$ orbitals in the laser-dressed ZnO should lead to the weakening of the Zn-O bond. We compared the ground-state electron density for the energy window around the bonding peak to the time-averaged laser-dressed electron density for the energy window around the new position of the corresponding peak. This comparison confirms the weakening of the bond (see the Supplementary Materials (SM) for details).

Summing up the discussion about the time-averaged PDOS, the broadening of the distributions and the reduction in the intensity of the prominent peaks can be attributed to a redistribution of the occupation probabilities between the valence and conduction states. We observe that the energies of the antibonding orbitals are shifted by a common value when laser dressed. The bonding orbitals do not experience the same energy shift, which can lead to a bond breaking.


\subsection{Time-dependent part of PDOS}

When studying a laser-dressed DOS, it is common to focus only on the time-averaged part of the DOS \cite{PhysRevLett.76.4576,enders2004electronic,lima2008effect,lima2009terahertz,oka2009photovoltaic,zhou2011optical,suarez2012radiation,calvo2012laser,PhysRevB.91.045439,PhysRevB.102.220301,PhysRevResearch.2.023274,Park2022,PhysRevB.108.235429,LIU2024415798,Zhan2024}. This provides a useful basis for interpreting the time-averaged properties of a laser-dressed system, such as its time-unresolved absorption spectra or conductivity. However, in recent years, subcycle-resolved experiments have become possible, enabling the dynamics of a system to be monitored during its interaction with a driving pulse \cite{wang2013observation,mahmood2016selective,sidiropoulos2021probing,graanas2022ultrafast, sidiropoulos2023enhanced,ito2023build,zhou2023pseudospin}. Our motivation for establishing the formalism for calculating the time-dependent part of the PDOS is that it can be useful for interpreting such experiments. In addition, it is a tool to gain insight into the electron dynamics during the interaction with light, as we will show.

Let us develop an understanding for the time-dependent part of DOS by considering a simple example. First, we observe that the DOS of the considered two-level system in \eq{eq_DOS2level} is time-independent. That is why we consider the following hypothetical Floquet state with an energy $E_h$ that would lead to a time-dependent term in the DOS
\begin{align}
\Psi_h(\br,t) = \frac{e^{-iE_{h}t}}{\sqrt{V}}&\lf(C_{10} \phi_1(\br) +C_{11} \phi_1(\br)e^{-i\Omega t}+ C_{21}  \phi_2(\br)e^{-i\Omega t}\rt).
\end{align}
The DOS that is due to only this Floquet state and its replicas is
\begin{align}
D_h(\omega,t)=&|C_{10} |^2\delta(\omega - E_h)+\lf(|C_{11}|^2+|C_{21}|^2\rt)\delta\lf(\omega - \lf[E_h+\Omega\rt]\rt)\nonumber\\
&+2\Re\lf(C^*_{11}C_{10} e^{i\Omega t} \rt)\delta\lf(\omega - \lf[E_h+\Omega/2\rt]\rt).\label{Eq_hypotheticalDOS}
\end{align}
The time-dependent term has the intensity $2\Re\lf(C^*_{11}C_{10} e^{i\Omega t}\rt)$. At the same time, the occupation probability of the state $|1\ra$ is equal to $|C_{10} |^2+2\Re\lf(C^*_{11}C_{10} e^{i\Omega t} \rt)$. This means that the time-dependent part of the DOS is related to the time-dependent part of the occupation probability of an electronic state. This conclusion can be generalized by considering occupation probabilities of electronic states $|n\k\ra$ of a system occupying a Floquet state $|\Phi_{i\mathbf k}\ra$. The time-dependent part of the DOS would contain the interference terms of these probabilities. Thus, the time-dependent part of the DOS arises due to interference terms in occupation probabilities of electronic states. 

The spectral position of the time-dependent term in the DOS in \eq{Eq_hypotheticalDOS} is in the middle between the energy of the main peak at $E_h$ and its side peak at $E_h+\Omega$. In general, the interference terms are centred at $E_{i\mathbf{k}}+(\mu'+\mu/2)\Omega$ for $\mu\neq0$ according to \eq{Eq_Dalm_Amp}, which is either in the middle between a main peak and its side peaks, or in the middle between side peaks. We observed the analogous behavior of a signal due to an optical dressing in our study of subcycle-resolved x-ray diffraction \cite{popova2024atomic}, where we theoretically investigated time-, momentum- and energy-resolved x-ray signals from a laser-dressed material. We showed that the time-dependent signal in this process consists of peaks with time-independent amplitudes and interference peaks with time-dependent amplitudes. The peaks with time-independent amplitudes are shifted by $n\Omega$ from a ground-state signal, where $n$ is an integer. The interference peaks are centered right in the middle between the spectral positions of the time-independent peaks, {\it i.e.}~are shifted by $n\Omega/2$. Thus, the behavior of the laser-dressed DOS is related to the behavior of time-dependent, subcycle-resolved spectra of laser-driven systems.

Colormaps in Figs.~\ref{Fig_LDPDOS}(d)-(f) and \ref{Fig_LDPDOS}(j)-(l) show the time-dependent part of the PDOS. The total PDOS is the sum of the time-dependent part and the time-averaged part shown on the corresponding panels to the left of the colormaps [Figs.~\ref{Fig_LDPDOS}(a)- (c) and \ref{Fig_LDPDOS}(g)-(i)]. The total PDOS is positive at any given energy.

We observe that the angle-resolved PDOS $\mathcal D^{\mathrm{O}}_{p_x}$, $\mathcal D^{\mathrm{O}}_{p_z}$, $\mathcal D^{\mathrm{Zn}}_{z^2}$ and $\mathcal D^{\mathrm{Zn}}_{r^2-z^2}$ in Figs.~\ref{Fig_LDPDOS}(d), (e), (j) and (k) have some regions with pronounced oscillations at the frequency $\Omega$ and other regions with pronounced oscillations at the frequency $2\Omega$. However, the angle-averaged PDOS $\mathcal D^{\mathrm{O}}_{p}$ in Fig.~\ref{Fig_LDPDOS}(f) and $\mathcal D^{\mathrm{Zn}}_{d}$ in Fig.~\ref{Fig_LDPDOS}(l) predominantly oscillate at the frequency $2\Omega$. The absence of $\Omega$ oscillations in angle-averaged distributions indicates that charge oscillations of frequency $\Omega$ around zinc and oxygen cancel out after angle averaging.


Let us analyze the oscillations of the PDOS by comparing it to the oscillations of the laser-dressed electron density. In Ref.~\cite{popova2024microscopic}, we showed that the time-dependent density of a material with a time-reversal symmetry driven by an electromagnetic field with the electric field $\mathbf{E}(t)=\mathbf{F}_0 \sin\Omega t$ evolves in time as
\begin{align}
\rho(\br,t) = \rho_0(\br) &- \sum\limits_{\mu_{\rm odd}\geq 1} \rho_{ \mu_{\rm odd}}(\br)\sin(\mu_{\rm odd}\Omega t) \nonumber \\
&+ \sum\limits_{\mu_{\rm even}\geq 2} \rho_{ \mu_{\rm even}}(\br)\cos(\mu_{\rm even}\Omega t).
\label{Eq_TD_density}
\end{align}
Here, $\rho_0(\br)$ is the time-averaged electron density and $\rho_\mu$ are the real-valued amplitudes of the oscillations. A $\mu$th-order amplitude of the electron density leads to a $\mu$th-order oscillation amplitude of polarization. For example, $-\rho_1\sin(\Omega t)$ leads to the oscillation of polarization as $\mathbf P^{(1)}\sin(\Omega t)$. The electron density and the PDOS have the analogous time dependence. The $\mu$th-order density amplitudes $\rho_\mu(\br)$ are related to the $\mu$th-order amplitudes of PDOS $\mathcal{D}^{\alpha}_{lm, \mu}(\omega) $ and describe the $\mu$th-order microscopic optical response of a material.

\begin{figure*}[htbp]
	\centering
	\includegraphics[width = 0.9\textwidth]{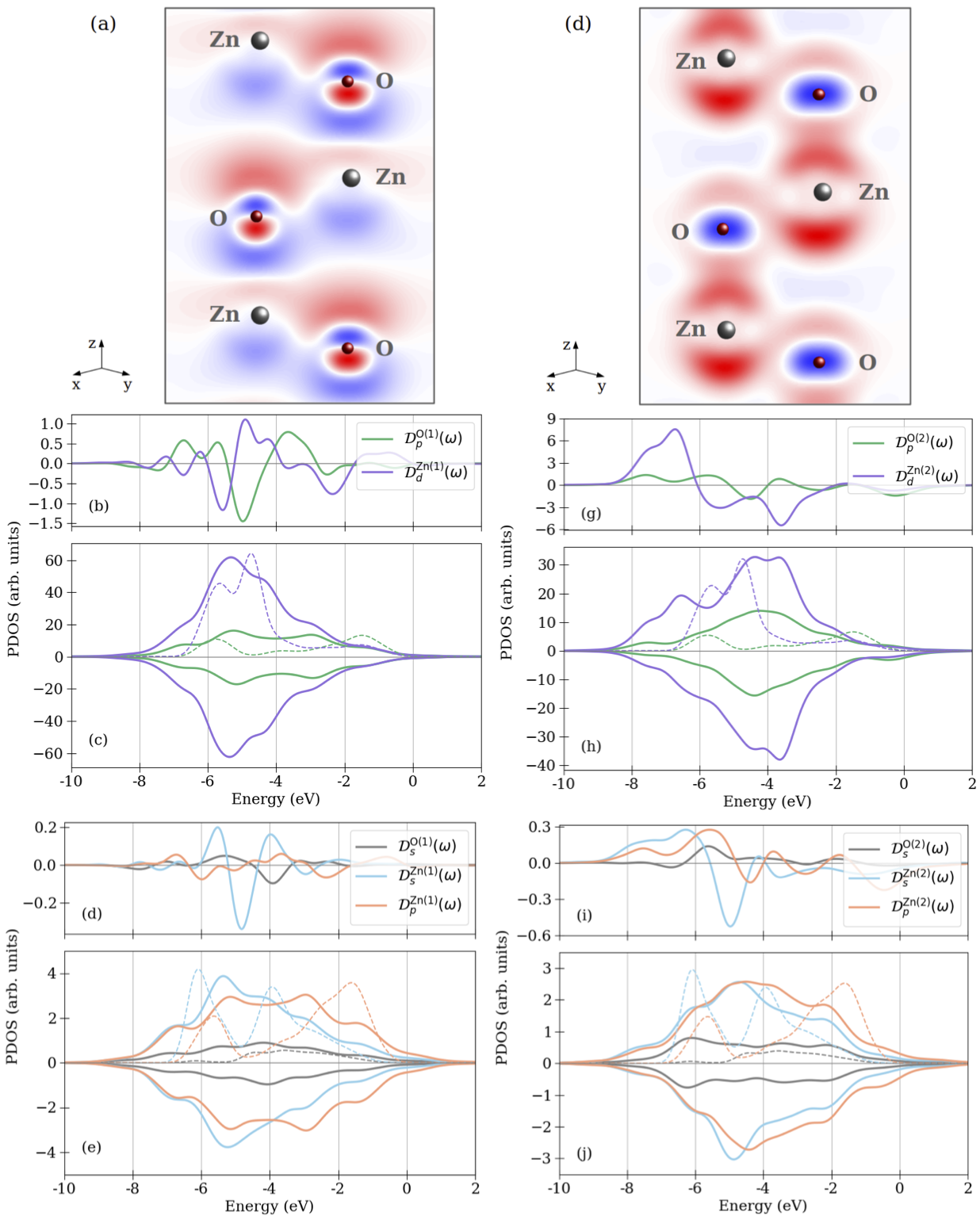}
	\caption{First- and second-order amplitudes of the electron density and the partial density of states (PDOS). (a) Two-dimensional cut through the first-order density amplitude $\rho_1(\br)$ in the plane shown in Fig.~\ref{fig:DOS}(a). The red and blue colors represent negative and positive charges, respectively. (b) First-order angle-averaged amplitudes $\mathcal{D}^{\rm O(1)}_{p}(\omega)$ and $\mathcal{D}^{\rm Zn(1)}_{d}(\omega)$. (c) Positive and negative contributions to the angle-averaged amplitudes in (b). (d) First-order amplitudes $\mathcal{D}^{\rm O(1)}_{s}(\omega)$, $\mathcal{D}^{\rm Zn(1)}_{s}(\omega)$, and $\mathcal{D}^{\rm Zn(1)}_{p}(\omega)$. (e) Positive and negative contributions to the amplitudes in (d). (f) Two-dimensional cut through the second-order density amplitude $\rho_2(\br)$ in the same plane as (a). (g) Second-order angle-averaged amplitudes $\mathcal{D}^{\rm O(2)}_{p}(\omega)$ and $\mathcal{D}^{\rm Zn(2)}_{d}(\omega)$. (h) Positive and negative contributions to the angle-averaged amplitudes in (g). (i) Second-order amplitudes $\mathcal{D}^{\rm O(2)}_{s}(\omega)$, $\mathcal{D}^{\rm Zn(2)}_{s}(\omega)$, and $\mathcal{D}^{\rm Zn(2)}_{p}(\omega)$. (j) Positive and negative contributions to the amplitudes in (i). The dotted lines in (c), (e), (h), and (j) in a corresponding color show the ground-state PDOS with reduced amplitudes for ease of comparison.}
	\label{Fig_cuts}
\end{figure*}

Figure \ref{Fig_cuts}(a) shows a two-dimensional (2D) cut through the first-order amplitude $\rho_1(\br)$ (see also Supplementary Fig.~S2). Charges rearrange around atoms and form alternating positively- and negatively-charged regions. Due to the interaction with the electric field of light, dipoles around oxygen and zinc are formed and give rise to polarization parallel to the electric field. Despite considerable charge distributions, the net charges around Zn and O are nearly neutral. This explains why the $\Omega$ oscillations are unnoticeable in the angle-averaged PDOS, but are pronounced in the angle-resolved PDOS. 

The optically-induced charge distributions $\rho_\mu$ are neutral, when averaged over the unit cell \cite{popova2024microscopic}. The positively- and negatively-charged regions switch their signs after half a cycle. Their magnitudes are maximal at $1/4$ and $3/4$ of the optical period, which is also when the first-order oscillations of the angle-resolved PDOS have maxima and minima Figs.~\ref{Fig_LDPDOS}(d), (e), (j) and (k). 

In addition to the evident effect of the occurrence of dipoles around the atoms, we observe that bonds between positively-charged regions around Zn and O as well as bonds between negatively-charged regions around Zn and O are formed. We analyze the structure of the bonds using the first-order amplitudes of PDOS $\mathcal{D}^{\alpha(1)}_{lm}(\omega) $, {\it cf.}~\eq{Eq_damplitudes}. Figure \ref{Fig_cuts}(b) shows angle-averaged amplitudes $\mathcal{D}^{\rm O(1)}_{p}$ and $\mathcal{D}^{\rm Zn(1)}_{d}$. They change sign several times over the energy scale because of competing positive and negative contributions to the total distributions, which makes them difficult to analyze. To facilitate the analysis, we disentangle the positive and negative contributions to the PDOS by separately summing positive and negative terms in \eq{Eq_Dalm_Amp} during the calculations. The resulting contributions are shown in Fig.~\ref{Fig_cuts}(c). The sum of the distributions in Fig.~\ref{Fig_cuts}(c) gives the PDOS in Fig.~\ref{Fig_cuts}(b). The positive contributions to the amplitudes of the PDOS correspond to a negative charge and the negative contributions correspond to a positive charge.

Looking at the disentangled contributions to the PDOS, we can analyze how different orbitals contribute to the formation of optically-induced charge distributions due to the first-order optical response. The positive and negative distributions in Figs.~\ref{Fig_cuts}(c) are almost identical, which reflects the similar shapes of the positive and negative regions in the first-order density amplitude.

The laser-dressed first-order Zn $d$-projected PDOS in Fig.~\ref{fig:DOS}(b) consists of five peaks forming one large peak. According to the crystal field theory, the $e_g-t_{2g}$ splitting in the case of ZnO is due to charges localized on oxygen atoms. Although the net charge around the oxygen is nearly neutral in the first-order response, there is still a small imbalance between the charges on Zn and O, see Fig.~\ref{Fig_cuts}(a). The net charge around Zn is positive and the net charge around O is negative. Because of this small imbalance, there is still a small $e_g-t_{2g}$ splitting in the distribution. The electric field of the light polarized along the $z$ axis breaks the symmetry of the tetrahedral field. This leads to the additional splitting of the $e_g$ and $t_{2g}$ states, resulting in the five peaks of the $d$-projected PDOS. The positions of the peaks in the O $p$-projected PDOS and in the Zn $d$-projected PDOS continue to align. This indicates that the O $p$ - Zn $d$ hybridized orbitals respond synchronously to the optical drive in the first-order optical response.

In the ground state, the O $s$-, Zn $s$-, and Zn $p$-projected PDOS are approximately one order of magnitude smaller than the O p- and Zn d-projected PDOS. Figure~\ref{Fig_cuts}(d) shows the first-order amplitude of the O $s$, Zn $s$, and Zn $p$-projected PDOS. Their weight relative to the Zn $d$- and O $p$-projected PDOS increases by about a factor of two in the first-order response. Dipole transitions from O $p$ and Zn $d$ states to O $s$ and Zn $p$ states explain the relative increase of the corresponding distributions. However, an analysis of dipole-allowed transitions alone is insufficient to explain transitions in a material with bonds. The relative increase of the Zn $s$-projected PDOS indicates that these states also contribute to the response via their contribution to hybridization.


Figure \ref{Fig_cuts}(g) shows the second-order density amplitude. In contrast to the first order, the positively- and negatively-charged regions of this charge distribution have a completely different shape. A positive charge localized on oxygen is formed. This positive charge is surrounded by a negative charge that repeats the shape of the positive charge in the vicinity of oxygen. Further away from oxygen, a strong bond to zinc along the $z$ axis is formed by a negative charge. Weak bonds with zinc atoms are also formed by a negative charge at the other positions. In addition, a delocalized positively-charged region is formed between the Zn and O atoms (see Supplementary Fig.~S3). The charge around the Zn atom is negative. Note that the second-order density amplitude oscillates with the frequency $2\Omega$. After a quarter of the optical period, the positively and negatively charged regions switch the signs.

The shapes of the positive and negative contributions to the second-order PDOS are also quite different as shown in Fig.~\ref{Fig_cuts}(h). As the consequence, their sum leads to a considerable total second-order O $p$- and Zn $d$-projected PDOS shown in Fig.~\ref{Fig_cuts}(g). It is overall much larger than the corresponding first-order angle-averaged PDOS. For this reason, the oscillations at the frequency $2\Omega$ dominate in the time evolution of the angle-averaged PDOS in Figs.~\ref{Fig_LDPDOS}(f) and (l).

The positive contribution of the Zn $d$-projected PDOS, corresponding to the negative charge, is split into two large peaks. This is caused by the inhomogeneous distribution of negative charges around Zn, see Fig.~\ref{Fig_cuts}(f). The orbitals localized on Zn are more strongly repelled by the negative charge around the oxygen in the $z$ direction than by the charges around the other oxygen atoms.


In contrast, the negative contribution of Zn $d$-projected PDOS contains only the right peak and does not have the strongly bound left peak. This indicates that the surrounding of the positively charge region formed by Zn $d$ orbitals is much more homogeneous compared to the surrounding of the negatively-charged regions. Consistently, the positive delocalized charge between the Zn and O atoms (see also Fig.~S3) is surrounded by an almost homogeneously distributed negative charge. This should be the reason why the splitting of the Zn $d$-projected PDOS is much smaller in the negative contribution than in the positive contribution.

The second-order amplitude of the O $p$-projected PDOS has a large peak formed by several peaks. The position of these peaks coincides with the position of the peaks in the second-order amplitude of the Zn $d$-projected PDOS, again indicating orbital hybridization. The peak has the maximum at larger binding energies for the positive charge than for the negative charge. The positive charge is more localized around oxygen than the negative charge, which explains the difference in binding energies.

The weight of O $s$-, Zn $s$- and Zn $p$-projected PDOS relative to Zn $d$- and O $p$-projected PDOS increase again in the second-order response, see Fig.~\ref{Fig_cuts}(j). The relative contributions of these states are about four times larger compared to the ground state. The increase of the role of Zn $s$-like states is due to two-photon transitions from Zn $d$ states.  The increase of the relative weights of O~$s$ and Zn~$p$-projected PDOS should be via bonds that involve these states. Additionally, it should be due to an increasing role of two-photon transitions from Zn $3p$-like states and O $2s$-like states to unoccupied states of the Zn $p$-like and O $s$-like character, correspondingly.

\section{Discussion}

In this study, we developed a computational framework for calculating a laser-dressed PDOS, which is a time-dependent PDOS of a material in the presence of periodic driving. Using a laser-driven ZnO as an example, we demonstrated that the laser-dressed PDOS is a useful tool for analysing laser-driven electron dynamics. We focused on the laser-dressed states that have evolved from initially occupied electronic states. Under the assumption of adiabatic switching, such laser-dressed states must become occupied and determine the laser-dressed electron dynamics.

In the first part of the article, we focused on the time-averaged part of the PDOS, which relates to the time-averaged occupation probability of electronic states. States close to the Fermi level shift towards higher binding energies more than states further away. The splitting between the bonding and antibonding states decreases, indicating a reduction in the overlap between the oxygen and zinc orbitals. The Zn $d$- and O $p$-projected bonding states shift by different amounts, which should also weaken the zinc-oxygen bonds.

We then considered the time-dependent part of the PDOS. Due to the electric field of the light, the electron density redistributes to form positively and negatively charged regions, oscillating with an optical period of $2\pi/\Omega$. These oscillations can be expanded in orders of the oscillation frequency $\Omega$, $2\Omega$, etc. At first order, the charge distributions mainly lead to the formation of dipoles around zinc and oxygen atoms. The dipoles centred on different atoms are bound together. To analyze the character of these distributions and bonds, we considered the amplitude of the PDOS oscillating at the frequency $\Omega$ and decomposed it into the positive and negative contributions corresponding to the negatively and positively charged regions, respectively. The structure of these distributions reveals that the Zn $d$- and O $p$-orbitals remain to be hybridized.

Charge redistribution due to the second-order optical response involves more than just the formation of dipoles \cite{popova2024microscopic}. Similarly, the structures of the contributions to the second-order amplitude of the PDOS corresponding to negative and positive charge differ considerably. There is a clear relationship between the features of these contributions and the charge distributions. For instance, significant splitting of the Zn d-projected PDOS occurs only when the charge around zinc is strongly repelled by the charge around oxygen in one direction.

For clarity, in our study we analyzed the amplitudes of the oscillations of the PDOS. To investigate the structure of the instantaneous bonds formed at a given time during the optical cycle, the same procedure and considerations should be applied to the laser-dressed PDOS at that time. We focused on the first- and second-order response in our study, but high-order electron oscillations can be studied with the same method. High-order amplitudes reveal electronic processes that lead to generation of corresponding high harmonics.

A PDOS is a well-known tool for analysing the electronic structure of materials in their ground state, providing insight into their bonding structure. The laser-dressed PDOS provides similar information about the structure of bonds that form the time-dependent electron density. It can serve as a new tool for analysing laser-driven electron dynamics. The information about instantaneous bonds provided by the laser-dressed PDOS can be used to trigger electronic processes in a controlled way.



\section{Methods}

\subsection{Computational details}

The one-body wave functions $\varphi_{n\mathbf{k}}(\mathbf{r})$ of the field-free Hamiltonian $H^{(0)}$, which describe the ground state of wurtzite ZnO, are calculated using the (L)APW$+$lo method implemented in the Exciting Code \cite{0}. We use the generalized gradient approximation functional of Perdew, Burke, and Ernzerhof (GGA-PBE) \cite{GGA-PBE}. We consider the $3d^{10}4s^2$ electrons of Zn and the $2s^2 2p^4$ electrons of O as valence electrons. We apply the scissors approximation \cite{Sciss} and correct the calculated band gap 0.77 eV to the experimentally-reported gap of around 3.4 eV \cite{ozgur2005comprehensive}.  Ref.~\cite{ZnO} has shown that the DOS of the conduction states calculated in the GW approximation agrees well with the one obtained in the DFT calculation using the GGA-PBE potential except for an energy shift. This validates the application of the scissor operator in cases where accurate band gap estimation is not required. Gaussian broadening of 0.27 eV is used.
	
We use the calculated Bloch wave functions as the basis functions to construct and diagonalize the Floquet-Bloch Hamiltonian as described in \cite{hsu2006floquet}. The number of blocks of the Floquet-Bloch matrix, $k$-points, and bands are increased in the computations until convergence of the Fourier components of the optically induced PDOS is reached. The laser-induced changes are converged for a $16 \times 16 \times 10$ $\mathbf{k}$-point grid, 150 conduction states, and $2\mu_{\text{max}}+1 = 121$ blocks of the Floquet Hamiltonian. In the calculations of the laser-dressed PDOS, we consider the states up to $l = 2$ in the case of oxygen and up to $l = 3$ in the case of zinc. States with higher $l$ are not included, because their energies are more than 30 eV above the band gap and thus cannot be relevant to the laser-induced dynamics (see Fig.~S1).

The optically induced charge distributions were computed using the formalism developed in Ref.~\cite{popova2024microscopic}. The $\mu$th-order density amplitudes, $\rho_{\mu}(\mathbf{r})$, entering Eq.~\eqref{Eq_TD_density}, were calculated considering only the plane wave part of the one-body Kohn-Sham wave functions, $\varphi_{n\mathbf{k}}(\mathbf{r})$. The calculations were performed on a single unit cell of wurtzite ZnO using a spatial grid of $50 \times 50 \times 80$ points.

\section{Competing Interests}

The Authors declare no Competing Financial or Non-Financial Interests.

\section{Author contributions}

T.B.~derived equations, wrote the code and performed calculations. D. P.-G.~supervised the study. Both authors analyzed the data and wrote the paper.

\section{Acknowledgments}
This work is funded by the Cluster of Excellence ``CUI: Advanced Imaging of Matter'' of the Deutsche Forschungsgemeinschaft (DFG)
-- EXC 2056 -- project ID 390715994. Daria Popova-Gorelova acknowledges the funding from the Volkswagen Foundation through a Freigeist Fellowship, grant number 96 237. The authors would like to thank Dmitry Tumakov for his valuable comments.

\bibliography{lit}

\end{document}